\documentstyle[aps,prd]{revtex}
\begin{document}

%%%%%%%%%%%%%%%%%%%%%%%%%%%%%%%%%%%%%%%%%%%%%%%%%%%%%%%%%%%%%%%%%%%%%%
\def \bq {\begin{equation}}
\def \eq {\end{equation}}
\def \pd {\partial}
\def \lb {\label}
%%%%%%%%%%%%%%%%%%%%%%%%%%%%%%%%%%%%%%%%%%%%%%%%%%%%%%%%%%%%%%%%%%%%%%

\title{Backreaction effects of dissipation in
neutrino decoupling}

\author{
Roy Maartens$^1$\thanks{email: roy.maartens@port.ac.uk}
and Josep Triginer$^2$\thanks{email: pep@ulises.uab.es}
}

\address{
$^1$School of Computer Science and Mathematics, Portsmouth
University, Portsmouth~PO1~2EG, England
}
\address{
$^2$Department of Physics, Autonomous University of
Barcelona, 08193~Bellaterra, Spain
}

\date{\today}

\maketitle

\begin{abstract}

Dissipative effects during neutrino decoupling in the early
universe create a small backreaction on the Hubble rate,
and lead to a small rise in temperature and entropy.
We use a simplified thermo-hydrodynamic model, which provides
a causal approximation to kinetic theory, in order to estimate
the backreaction effects and the entropy production.

\end{abstract}


\section{Introduction}

Non-equilibrium processes in the early universe are typically
associated with dynamical transitions or particle decouplings. In
the case of neutrino decoupling, the standard approach is to treat
the process as adiabatic (see e.g. \cite{w}). The small
non-equilibrium effects are thus usually neglected, which provides
a reasonable approximation. However, given the increasing accuracy
of cosmological observations and theoretical modeling, it is
worthwhile revisiting the standard equilibrium models of processes
such as neutrino decoupling, in order to see whether
non-equilibrium corrections can lead to observable consequences.
Recently, non-equilibrium corrections in neutrino decoupling have
been calculated in a number of papers, using complicated kinetic
theory and numerical computations (see \cite{Dolgova} for a short
review). The corrections are very small, as expected. For example,
in \cite{Herrera,Mitra,fks} it was found that non-equilibrium
effects lead to a small change in the decoupling temperature for
neutrinos. Spectral distortions have also been analyzed
\cite{Dolgovb}, showing the remarkable fact that they amount to as
much as 1\% or more for the higher-energy side of the spectrum.
Although these corrections in the spectrum, energy density and
temperature of the neutrino component have hardly any effect on
primordial helium synthesis, yielding a change in the mass
fraction of $\sim 10^{-4}$, they can lead to other effects that
may be observable. Thus it is shown that the non-equilibrium
increase in neutrino temperature, which leads to an extra
injection of energy into the photon spectrum, leads to a shift of
equilibrium epoch between matter and radiation which, in turn,
modifies the angular spectrum of fluctuations of the cosmic
microwave background radiation \cite{Gnedin,Dolgovc}.

Despite the accuracy of these models in obtaining corrections to
the decoupling temperature and distribution function due to
non-equilibrium effects, they still make use of the standard
Friedman equations for a perfect (i.e non-dissipative) fluid. This
leads to the physically inconsistent situation in which, say, the
energy density and expansion evolve in time like a radiative fluid
in equilibrium. One expects that small distortions in the particle
equilibrium distribution function should be reflected in the
macroscopic (i.e fluid) description, as given by the stress-energy
tensor, by adding a bulk viscous pressure to the equilibrium one.
Here we consider an alternative thermo-hydrodynamic model of
dissipative effects in neutrino decoupling, simple enough to
produce analytic solutions for the backreaction effects on the
universal scale factor, and estimates for the entropy production
due to dissipation. As explained above these effects are not the
focus of recent papers, which use sophisticated kinetic theory
models focusing on the neutrino temperature. Our simplified
approach cannot compete with these models for accuracy and
completeness, but it has the advantage of simplicity, allowing for
a qualitative understanding of effects not previously investigated
in detail. A similar approach has previously been developed in
\cite{zpm} to the reheating era that follows inflation.

The thermo-hydrodynamic model is based on an approximation to
kinetic theory which respects relativistic causality. This
approximation is the Grad moment method, leading to the causal
thermodynamics of Israel and Stewart \cite{is} in the hydrodynamic
regime (see also \cite{p} for an alternative but equivalent
approach). This causal theory is a generalization of the more
commonly used relativistic Navier-Stokes-Fourier theory. The
latter, due to Eckart \cite{e}, may be derived via the
Chapman-Enskog approximation in kinetic theory. The resulting
theory is quasi-stationary and noncausal, and suffers from the
pathologies of infinite wavefront speeds and instability of all
equilibrium states \cite{HisLin}. The main new ingredient in the
causal transport equations is a transient term which contains the
relaxation time. Our simple model is based on a one-component
fluid. In \cite{Herrerab}, relaxation time processes are
incorporated in a two-fluid model. In this setting, electrons and
positrons on the one side and neutrinos and antineutrinos on the
other side, are found to be in two different equilibrium states
with slightly different temperatures. The system evolves towards a
state of thermal equilibrium in a characteristic relaxation time.

Dissipative effects in the decoupling of a given species of
particles arise from the growing mean free path of the decoupling
particles in their weakening interaction with the cosmic fluid.
Eventually the mean collision time exceeds the gravitational
expansion time, and decoupling is complete. A hydrodynamic model
may be used to cover the early stages of the decoupling process,
but it will eventually break down when the mean collision time
becomes large enough \cite{mt1}.

In the conditions prevailing at the time of neutrino decoupling,
it is reasonable to neglect sub-horizon metric fluctuations and
treat the spacetime as a Friedmann model. (The incorporation of
perturbations in our model would use the covariant formalism for
dissipative fluids developed in \cite{mt2}.) The dynamical effects
of spatial curvature and any surviving vacuum energy will be
negligible, so that we can reasonably assume a spatially flat
geometry. Furthermore, we assume that the average 4-velocities of
the neutrinos (regarded as massless) and of the
photon-electron-positron gas are the same. With all these
assumptions, only scalar dissipation is possible. Dissipation
during neutrino decoupling arises because the falling temperature
lowers the interaction rate with leptons as the lepton mass can no
longer be ignored relative to the thermal energy. Thus dissipation
is directly reflected in a deviation of the equation of state from
the thermalized radiation form $p={1\over3}\rho$. Within a
hydrodynamic one-fluid model, such dissipation is described via
bulk viscosity, which vanishes in the $p={1\over3}\rho$ limit, but
is nonzero otherwise. We will use the full (i.e. non-truncated)
version of the causal transport equation for bulk stress.

\section{Causal transport equation for bulk stress}

The particle number 4-current and the energy-momentum tensor are
\[
N^a=nu^a\,,~ T^{ab}=\rho u^au^b+(p+\Pi)h^{ab}\,,
\]
where $\rho$ is the energy density, $p$ is the equilibrium
(hydrostatic) pressure, $n$ is the particle number density, $\Pi$
is the bulk viscous pressure, and $h^{ab}=g^{ab}+ u^au^b$ is the
projector into the comoving instantaneous rest space. Particle and
energy-momentum conservation
\[
\nabla_aN^a=0\,,~\nabla_bT^{ab}=0\,,
\]
lead to the equations
\begin{eqnarray}
&& \dot{n}+3Hn=0 \,,\lb{3a}\\
&& \dot{\rho}+3H(\rho+p+\Pi)=0 \,,\lb{3b}
\end{eqnarray}
where $H$ is the Hubble expansion rate. The specific entropy $s$
and the temperature $T$ are related via the Gibbs equation \bq
nTds=d\rho-\frac{\rho+p}{n}dn\,. \lb{5} \eq Then it follows that
\bq nT\dot{s} =-3H\Pi   \,, \lb{ent}\eq where $\Pi$ is always
non-positive. The Grad moment approximation in kinetic theory (or
phenomenological arguments) leads to the full causal transport
equation \cite{is} for $\Pi$: \bq \tau\dot{\Pi}+\Pi=-3\zeta
H-{\textstyle{1\over2}}\tau\Pi\left[3H+ \frac{\dot{\tau}}{\tau}
-\frac{\dot{\zeta}}{\zeta}-\frac{\dot{T}}{T}\right] \,, \lb{6a}
\eq where $\tau$ is the relaxation time scale, which allows for
causal propagation of viscous signals, and $\zeta\leq 0$ is the
bulk viscous coefficient as given below. Quasi-stationary,
noncausal theories have $\tau=0$, which reduces the evolution
equation (\ref{6a}) to an algebraic equation $\Pi=-3\zeta H$. This
leads to instantaneous propagation of viscous signals. Note also
that the causal relaxational effects lead to a small increase in
the sound speed over its adiabatic value \cite{MaarRund}: \bq
c_{\rm s}^2~\rightarrow~c_{\rm s}^2+c_{\rm b}^2~\mbox{ where }
~c_{\rm b}^2={\zeta\over(\rho+p)\tau} \,. \lb{cb}\eq This result,
which is not well known, is derived in the appendix.

The approximation used in deriving the transport
equation (also in the quasi-stationary case) requires that
$|\Pi|\ll \rho$, which is reasonable for most dissipative
processes (see \cite{mm} for a nonlinear generalization
of the causal transport equation.)

Equation (\ref{6a}) as it stands is known as the full or
non-truncated transport equation for bulk viscous pressure
\cite{HisSal,Maartens95,z}. When the term containing the square
bracket on the right is neglected, we get the truncated equation
which is usually used. Under many conditions, truncation leads to
a reasonable approximation. We will use the full equation.

Taking $n$ and $\rho$ as independent
variables, the Gibbs equation (\ref{5})
leads to the integrability condition
\bq
n\left(\frac{\pd T}{\pd n}\right)_{\rho}+(\rho+p)\left(\frac{\pd T}
{\pd\rho}\right)_n=T\left(\frac{\pd p}{\pd\rho}\right)_n \,,
\lb{7}\eq
and together with the energy conservation equation (\ref{3b})
this gives the temperature evolution equation
\bq
\frac{\dot{T}}{T}=
-3H\left(\frac{\pd p}{\pd\rho}\right)_n-\frac{1}{T}\left(
\frac{\pd T}{\pd\rho}\right)_n3H\Pi \,.
\lb{8}
\eq
The first term on the right accounts for adiabatic cooling due to
expansion, whereas in the second term, viscosity
contributes to heating of the fluid
(note that $\Pi$ is always non-positive).

Using equations
(\ref{3a}) and (\ref{3b}), the Gibbs equation takes the form
\bq
n^2Tds=\left[\frac{n3H\Pi}{3H(\rho+p)+3H\Pi}\right]d\rho+(\rho+p)
\left(\frac{\pd n}{\pd p}\right)_{\rho}
\left[\frac{\dot{p}}{\dot{\rho}}\,d\rho-
dp\right] \,.
\lb{9}
\eq
As expected we learn from the
last equation that when the fluid is perfect
($\Pi=0$), the specific entropy is conserved
along the flow lines ($\dot{s}=0$).
Furthermore, if a barotropic equation of
state for $n$ holds, i.e. $n=n(\rho)$, then $ds=0$ so that $s$
is a universal constant, the same on all flow-lines, and
the fluid is called isentropic.\footnote{
The same reasoning applies
when the temperature is barotropic.
}
Yet, as Eq. (\ref{9})
shows, this is no longer true in the presence of dissipation, i.e. a
barotropic particle number density no longer forces $ds$ to vanish.

For simplicity, we assume the linear barotropic equation of state
\bq
p=(\gamma-1)\rho \,,
\lb{10}
\eq
where $\gamma$ is constant and
we are interested in the case $\gamma\approx {4\over3}$.
The adiabatic speed of sound
$c_{\rm s}$ is given by
\[
c_{\rm s}^2=\left(\frac{\pd p}{\pd\rho}\right)_s \,,
\]
which for a perfect fluid (either barotropic or not) becomes
\[
c_{\rm s}^2=\frac{\dot{p}}{\dot{\rho}} \,.
\]
When Eq. (\ref{10}) holds then $c_{\rm s}=\sqrt{\gamma-1}$.
Using Eq.
(\ref{10}) and the integrability condition (\ref{7}), we find
\bq
T=\rho^{(\gamma-1)/\gamma}F\left(\frac{\rho^{1/\gamma}}{n}\right)\,,
\lb{11}
\eq
where $F$ is an arbitrary function which satisfies $\dot{F}=0$. If
$T$ is barotropic, then $F$ is constant and we have a power-law form
with fixed exponent for the temperature \cite{MaarRund,MeniTri}
\bq
T\propto \rho^{(\gamma-1)/\gamma} \,.
\lb{12}
\eq
In the non-dissipative case, these barotropic equations for
$p$ and $T$ are compatible with the ideal gas law
\bq
p=nT \,,
\lb{13}
\eq
but in the presence of dissipation this is no longer true. In effect,
equations (\ref{10}), (\ref{12}) and (\ref{13}) imply
$n\propto\rho^{1/\gamma}$, i.e.
\[
\frac{\dot{n}}{n}=\frac{1}{\gamma}\frac{\dot{\rho}}{\rho} \,,
\]
which implies, by using Eq. (\ref{3b}), that $\Pi=0$.
We shall drop in the sequel a barotropic equation of state for
the temperature in favour of the more physically appealing equation
of state (\ref{13}) together the $\gamma$-law in (\ref{10}).

\section{Dissipation in neutrino decoupling}

A hydrodynamic approach in the expanding universe requires
a particle collision time $t_{\rm c}$ short enough to
adjust to the falling
temperature.
As the natural time-scale for the expanding universe is
$H^{-1}$, we have
\[
t_{\rm c}<H^{-1} \,.
\]
If $t_{\rm c}\ll H^{-1}$, then an equilibrium state
can in principle be
attained. Dissipative phenomena could play a prominent
role
for $t_{\rm c}\sim H^{-1}$.

We learn from kinetic theory that $t_{\rm c}$ is determined by
\bq
t_{\rm c}=\frac{1}{n\sigma v} \,,
\lb{tc}\eq
where $n$ is the number density of the target particles with which
the given species is interacting, $\sigma$ the cross-section and $v$
the mean relative speed of interacting particles.
For the decoupling of massless neutrinos in the early
universe, $v= 1$, the target number density is that of
electrons, and \cite{pad}
\[
\sigma \approx G_{_F}T^2 \,,
\]
where $G_{_F}$ is the Fermi
coupling constant. At the neutrino decoupling
temperature $T_{\rm d}$,
we have $m_{\rm e}/T_{\rm d}\approx {1\over2}$,
so that the rest mass energy $m_{\rm e}$ of electrons
starts
to become important.
Since the electron number
density in the radiation dominated era evolves as
$n_{\rm e}\propto a^{-3}$, where $a$ is the scale factor,
we have from Eq. (\ref{tc}) that
\bq
t_{\rm c}\propto \frac{a^3}{T^2} \,.
\lb{16}
\eq

Dissipation due to massless particles with long mean free path
in a hydrodynamic fluid is described by the radiative transfer
model.
The bulk viscous coefficient takes the form \cite{Weinberg}
\bq
\zeta=4rT^4\Gamma^2t_{\rm c},
\lb{17}
\eq
where $r$ is ${7\over8}$
times the radiation constant
and $\Gamma$ measures the deviation of $p/\rho$ from its
pure-radiation value:
\bq
\Gamma=\frac{1}{3}-\left(\frac{\pd p}{\pd\rho}\right)_n,
\lb{18}
\eq
where $p$ and $\rho$ refer to the pressure and energy density of
the radiation/matter mixture as a whole.
Since we assume the linear equation of state (\ref{10}), it
follows that $\Gamma$ is a perturbative constant parameter in our
simple model:
\[
\Gamma={\textstyle{4\over3}}-\gamma\ll 1 \,.
\]
The assumption that $\Gamma$ is constant relies on the assumption that
decoupling takes place rapidly. Since standard adiabatic treatments
of decoupling \cite{w} assume instantaneous decoupling, this
assumption should be a reasonable first approximation.

We may neglect the $-3\zeta H$ term on the right
of the transport equation (\ref{6a}), since it is $O(\Gamma^2)$.
Note that our simple model would thus break down in the
quasi-stationary Eckart theory, since it would immediately
lead to $\Pi=O(\Gamma^2)$.
The relaxation timescale $\tau$ in causal radiative transfer
\cite{ui}
is given by $\tau=t_{\rm c}$. The term $\dot{\zeta}/\zeta$
on the right of Eq. (\ref{6a}) becomes
\[
{\dot{\zeta}\over\zeta}=H+O(\Gamma) \,,
\]
on using equations (\ref{8}) and (\ref{16}).
The full
transport equation (\ref{6a}) becomes, to lowest order
\bq
\tau\dot{\Pi}+\Pi=-4\tau H\Pi \,.
\lb{19}
\eq
(We can think of the right hand side as an effective source term
relative to the truncated transport equation.)
We can rewrite this in the standard truncated form as
\bq
\tau_*\dot\Pi+\Pi=0 \,,
\lb{23}
\eq
where the effective relaxation time acquires an expansion correction:
\bq
\tau_*=\frac{\tau}{1+4\tau H} \,.
\lb{24}
\eq
The amount of reduction depends
on the size of $\tau=t_{\rm c}$ relative to
$H$. The
hydrodynamical description requires $\tau H<1$. If $\tau H\ll 1$,
then $\tau_*\approx\tau$. But if $\tau H$
is close to 1, the reduction could be significant.

The Friedmann equation
\bq
\rho=3H^2 \,,
\lb{26}
\eq
together with Eq. (\ref{3b}) leads to
\bq
\Pi=-2\dot{H}-(4-3\Gamma) H^2 \,.
\lb{27}
\eq
On using equation (\ref{27}) we get from (\ref{19}) the evolution
equation for $H$
\bq
\ddot H+H\dot{H}(8-3\Gamma+N)+H^3(2-\textstyle{3\over 2}\Gamma)(N+4)=0,
\lb{28}
\eq
where
\bq
N=(\tau H)^{-1} \,,
\lb{29}
\eq
which
is of the order of the number of interactions in an expansion
time.
Now, from equations
 (\ref{10}), (\ref{13}), (\ref{16}) and (\ref{29}) we have
\bq
N=\left(\frac{Ha}{H_{\rm d}a_{\rm d}}\right)^3 \,,
\lb{30}
\eq
where the expression $n\propto a^{-3}$ has been used and
$a_{\rm d}$ and $H_{\rm d}=H(a_{\rm d})$
are the values at which $N=1$, so that $a_{\rm d}$ is determined
by the equation
\bq
t_{\rm c}(a_{\rm d})H(a_{\rm d})=1 \,.
\lb{dec}\eq
Changing the independent variable
to the scale factor $a$, developing equation (\ref{28})
and collecting the previous results, yields
\begin{eqnarray}
a^2HH''+a^2H'^{2}+aHH'\left[9-3\Gamma
+\left(\frac{Ha}{H_{\rm d}a_{\rm d}}\right)^3\right]&& \nonumber\\
{}+\left(2-
{\textstyle{3\over2}}\Gamma\right)
H^2\left[4 +\left(\frac{Ha}{H_{\rm d}a_{\rm d}}\right)^3
\right]=0 \,,&&
\lb{32}
\end{eqnarray}
where a prime denotes $d/da$.
We expand $H$ as
\bq
H=\bar H+\delta H~\mbox{ where }~ \delta H=\Gamma h+O(\Gamma^2)
\,.
\lb{33}
\eq
The equilibrium Hubble rate $\bar{H}$
corresponds to the thermalized radiation
state $p={1\over3}\rho$, so that $\Gamma=0$, and
Eq. (\ref{33}) becomes
\[
a^2\bar{H}\bar{H}''+a^2\bar{H}'^2+9a\bar{H}\bar{H}'+8\bar{H}^2
+\left(a\bar{H}\bar{H}'+2\bar{H}^2\right)\left({\bar{H}a\over\bar{H}_{\rm
d}a_{\rm d}}\right)=0\,.
\]
The unique power-law solution is
the well-known perfect radiative solution
\bq
\bar{H}=H_0\left({a_0\over a}\right)^2={1\over2t}\,,
\lb{29a}\eq
where $a_0$ marks the start of the
dissipative decoupling process, so that $H=\bar{H}$ for $a<a_0$.

Substituting Eq. (\ref{33}) into (\ref{32}) and using the fact that
\[
{H_0a_0\over H_{\rm d}a_{\rm d}}={a_{\rm d}\over a_0}+O(\Gamma)\,,
\]
we find that to $O(\Gamma)$:
\bq
a^2h''+a\left[5+\left({a_{\rm d}\over a}\right)^3\right]h'+
\left[4+2\left({a_{\rm d}\over a}\right)^3\right]h={\textstyle{3\over2}}
H_0\left({a_0\over a_{\rm d}}\right)^2\left({a_{\rm d}\over a}\right)^5 \,.
\lb{35}
\eq
Defining $\alpha=a/a_{\rm d}$, we can rewrite this as
\bq
{d^2h\over d\alpha^2}+\left[{5\over
\alpha}+{1\over\alpha^4}\right]{d
h\over d\alpha}+\left[{4\over\alpha^2}+{2\over\alpha^5}\right]h=
\left({\textstyle{3\over2}}H_0\alpha_0^2\right)\,{1\over\alpha^7}\,.
\lb{35'}\eq
Now we use the following general result \cite{k}: if $\varphi$ is
a solution of
\[
y''+f(x)y'+g(x)y=k(x)
\]
when $k=0$, then the general solution is
\[
y=C_1\varphi+C_2\varphi\int{dx\over\varphi^2E}
+\varphi\int{1\over\varphi^2E} \left(\int\varphi Ekdx\right)d
x\,,
\]
where $E=\exp\int fdx$. By inspection, a solution of the
homogeneous equation (\ref{35'}) is $1/\alpha^2$. It follows that
the general solution is
\bq
h(a)=H_0\left({a_0\over a}\right)^2
\left\{c_1+c_2{\rm
Ei}\left[{1\over3}\left({a_{\rm d}\over
a}\right)^3\right]+{3\over2}\ln\left({a\over a_{\rm
d}}\right)\right\}\,,
\lb{36}
\eq
where $c_1$ and $c_2$ are
arbitrary integration constants and Ei is the exponential-integral
function \cite{gr}
\[
{\rm Ei}(x)\equiv\int_{-\infty}^x{e^v\over v}dv={\cal C}+\ln
x+\sum_{k=1}^\infty{x^k\over k!k}\,,
\]
with ${\cal C}$ denoting Euler's constant.

By equations (\ref{27}) and (\ref{36}), the bulk stress to first order is
\bq
\Pi=(3\bar{H}^2-4\bar{H}h-2h'Ha)\Gamma,
\lb{37}
\eq
This expression holds for $a>a_0$, where $a_0$ marks the onset of
dissipative evolution.
Thereafter, the bulk stress decays
according to the causal law (\ref{23}).
In order to relate the constants $c_1$ and $c_2$,
we require,
according to standard matching conditions, that $H$ is
continuous. Thus $h(a_0)=0$, which fixes $c_1$:
\bq
c_1=-c_2{\rm Ei}\left[{1\over3}\left({a_{\rm d}\over a_0}\right)^3\right]
-{3\over2}\ln\left({a_0\over a_{\rm d}}\right) \,.
\lb{38}
\eq
Thus, using Eq. (\ref{36}), we see that
the backreaction of the dissipative decoupling
process on the expansion of the universe is given by
\begin{eqnarray}
\delta H &=& \bar{H}\left\{c_2\left(
{\rm Ei}\left[{1\over3}\left({a_{\rm d}\over a}\right)^3\right]
-{\rm Ei}\left[{1\over3}\left({a_{\rm d}\over
a_0}\right)^3\right]\right)
+{3\over2}\ln\left({a\over a_0}\right)\right\}\Gamma\nonumber\\
{}&&{}+O(\Gamma^2)\,.
\lb{39}
\end{eqnarray}
Substituting Eq. (\ref{38}) into Eq. (\ref{37}), we find that the
bulk stress becomes
\bq
\Pi =\bar{\rho}\left\{2c_2 \exp
\left[{1\over3}\left({a_{\rm d}\over
a}\right)^3\right]\right\}
\Gamma +O(\Gamma^2)\,,
\lb{39'}
\eq
where $\bar{\rho}=3\bar{H}^2$ is the equilibrium energy density.
Since $\Pi<0$, we require $c_2<0$. Below we find a prescription
for $c_2$ in terms of physical parameters.

\section{Conclusion}

In order to complete the model, we need to determine the remaining
arbitrary constant $c_2$ in terms of physical parameters.
A rough estimate, which is consistent
with the simplicity of the model, arises as follows. We estimate
the duration of the dissipative process as
\bq
\Delta a\approx a_{\rm d}-a_0\,,
\lb{42}\eq
i.e. we assume that the process ends at $a_{\rm d}$. Then by Eqs.
(\ref{8}) and (\ref{13}), the fractional viscous rise in temperature due to
decoupling is approximately
\bq
{\Delta T\over T}\approx -{\Pi(a_0)\over\bar{\rho}(a_0)}\,
{\Delta a\over a_0}\,.
\lb{40}\eq
We can consider the fractional temperature increase
as an input
from previous kinetic-theory investigations (as
described in the introduction), which
typically predict it to be $O(10^{-3})$.\footnote{
Note that this small temperature increase is due to dissipative
heating, and is not to be confused with the larger temperature
increase arising from electron-positron annihilation, which occurs after
neutrino decoupling. Our model does not consider the annihilation
process.
}
Then equations (\ref{39'})--(\ref{40}) and (\ref{37}) allow us to
estimate the constant $c_2$ in terms of the physical parameters $a_{\rm d}/a_0$,
$\Delta T/T$ and $\Gamma$ as:
\bq
c_2\Gamma\approx-{1\over2}\,{\Delta T\over
T}\,\left\{{\exp\left[-{1\over3}\left({a_d\over
a_0}\right)^3\right]\over \left({a_d\over
a_0}\right)-1}\right\}\,.
\lb{40'}\eq

Finally, we can also estimate the entropy production due to
decoupling. By Eqs.
 (\ref{ent}) and (\ref{40}), the viscous increase in entropy
per particle is approximately
\bq
\Delta s\approx 3{\Delta T\over T}\,.
\lb{43}\eq

Our model describes the response of the
cosmic fluid to a bulk stress, which is a very simple
thermo-hydrodynamic approximation to more realistic kinetic theory
models of neutrino decoupling,
but which nevertheless accommodates the dissipative effects
and respects relativistic causality.
The simplicity of our model allows us to derive analytic forms
for the dynamical quantities and the backreaction effects, but
it does not incorporate a mechanism for bringing the dissipative
process to an end.

\[ \]
{\bf Acknowledgements:}

This work was partially supported by a European Science Exchange
Programme grant.

%\newpage

\appendix

\section{Characteristic velocities for bulk viscous perturbations}

Following \cite{MaarRund},
we derive equation (\ref{cb}) for the dissipative
contribution to the sound speed.
The full
analysis of the causality and stability of the Israel-Stewart theory
was performed in a series of papers
by Hiscock and Salmonson \cite{HisLin,HisLin2}.
They showed that both issues are closely related and obtained
general expressions for the characteristic velocities for dissipative
perturbations.
Here we extract from their general
expressions specific results for the
case in which only bulk viscosity is
present.

The purely bulk viscous case stems from the general expressions of
\cite{HisLin}
by setting all the coefficients coupled to heat flux and shear
viscosity to zero.
This yields for the speed of propagating transverse modes
\[
v_{_T}^2=
\frac{(\rho+p)\alpha_1^2+2\alpha_1+\beta_1}{2\beta_2
[\beta_1(\rho+p)-1]}
\rightarrow 0 \,,
\]
which is what one
expects for scalar sound-wave perturbations. Equation
(128)
of \cite{HisLin} governing the speed $v=v_{_L}$
of propagating longitudinal
modes becomes, on dividing by $\beta_0\beta_2$ and setting
$\alpha_0=\alpha_1=0$,
\begin{eqnarray*}
&&[\beta_1(\rho+p)-1]v^4\\
&&+\left[\frac{2n}{T}\left(\frac{\partial T}
{\partial n}\right)_s-\frac{(\rho+p)}{nT^2}\left(\frac{\partial T}
{\partial s}\right)_n-\beta_1\left\{(\rho+p)\left(\frac{\partial p}
{\partial\rho}\right)_s+\frac{1}{\beta_0}\right\}\right]v^2\\
&&+\frac{1}{nT^2}\left(
\frac{\partial T}{\partial s}\right)_n\left[(\rho+p)
\left(\frac{\partial p}{\partial\rho}\right)_s
+\frac{1}{\beta_0}\right]-
\left[\frac{n}{T}\left(\frac{\partial T}
{\partial n}\right)_s\right]^2=0 \,.
\end{eqnarray*}
Dividing by $\beta_1$ and taking $\beta_1\rightarrow\infty$, we have
\bq
v^2=\left(\frac{\partial p}{\partial\rho}\right)_s
+\frac{1}{(\rho+p)\beta_0}.
\lb{182}
\eq
The first term on the right
is the adiabatic contribution $c_{\rm s}^2$ to $v^2$,
and the second term is the dissipative contribution
$c_{\rm b}^2$, which, requiring $v^2\leq 1$, leads to
\bq
c_{\rm b}^2\equiv\frac{\zeta}{(\rho+p)\tau}\leq1-c_{\rm s}^2 \,.
\lb{183}
\eq
We also learn from \cite{HisLin} that causality and stability require
\bq
\Omega_3(\lambda)\equiv(\rho+p)\left\{1-\lambda^2\left[\left(
\frac{\partial p}
{\partial\rho}\right)_s+\frac{1}{(\rho+p)\beta_0}\right]\right\}
\leq 0 \,,
\lb{184}
\eq
for all $\lambda$ such that $0\leq\lambda\leq 1$. This condition
is seen to
be hold on account of the inequality (\ref{183}).

The expression for $c_{\rm b}$
refines and corrects the statement in \cite{Belinski} (the
first paper to apply causal bulk viscosity in cosmology) that
$\zeta/\rho\tau=1$ is required by causality.

\end{document}